\documentclass[12pt,twoside]{article}
\usepackage{fleqn,espcrc1,epsfig} %,floatfig}
\newcommand{\eg}{{\it e.g.\ }}
\newcommand{\ie}{{\it i.e.\ }}

\newcommand{\Pma}{I\!\!P}
\def\lsim{\mathrel{\rlap{\lower4pt\hbox{\hskip1pt$\sim$}}
    \raise1pt\hbox{$<$}}}         %less than or approx. symbol
\def\gsim{\mathrel{\rlap{\lower4pt\hbox{\hskip1pt$\sim$}}
    \raise1pt\hbox{$>$}}}         %greater than or approx. symbol

\title{Soft Colour Interactions in Non-perturbative QCD\thanks{Contribution to 
PANIC 99 conference proceedings}}

\author{G.\ Ingelman$^{ab}$\thanks{~ingelman@tsl.uu.se \hspace*{5mm} 
                                    www3.tsl.uu.se/thep/},  
        A.\ Edin$^b$, R.\ Enberg$^a$, J.\ Rathsman$^c$, N.\ Timneanu$^a$\\ ~~ \\
$^a$~Dept.\ of Radiation Sciences, Uppsala University, Box 535, S-751 21 Uppsala, Sweden\\
$^b$~DESY, Notkestrasse 85, D-22603 Hamburg, Germany\\
$^c$~SLAC, Stanford, California 94309, USA\\
}

\begin{document}

%\initfloatingfigs
\maketitle
\vspace*{-75mm}
\noindent
TSL/ISV-99-0222\\
October 1999
\vspace*{65mm}

\begin{abstract}
Improved understanding of non-perturbative QCD dynamics can be obtained 
in terms of soft colour exchange models. 
Their essence is the variation of colour string-field topologies giving a 
unified description of final states in high energy interactions. 
In particular, both events with and without large rapidity gaps are 
obtained in agreement with data from $ep$ at HERA and $p\bar{p}$ at the 
Tevatron, where also the surprisingly large production rate of high-$p_\perp$ 
charmonium and bottomonium is reproduced. 
\end{abstract}

\vspace{15mm}
Strong interaction processes at small (`soft') momentum transfers belong to the 
realm of non-perturbative QCD, which is a major unsolved problem in particle and
nuclear physics. High energy particle collisions involving a `hard' scale, 
\ie a large momentum transfer, has the advantage of providing a well
defined parton level process which is calculable in perturbative QCD (pQCD). 
The soft effects (\eg hadronisation) in such hard scattering events can 
therefore be investigated based on an understood parton level process. 

This hard-soft interplay is the basis for the topical research field of 
diffractive hard scattering \cite{IS,StCroix}. Diffractive events are 
characterised by having a rapidity gap, \ie a large region of rapidity
(or polar angle) without any particles. The rapidity gap connects to the 
soft part of the event and therefore non-perturbative effects on a long 
space-time scale are important. 

In order to better understand non-perturbative dynamics and provide a unified 
description of all final states, we have developed new models. 
These models are added to Monte Carlo generators 
({\sc Lepto} \cite{Lepto} for $ep$ and 
{\sc Pythia} \cite{Pythia} for $p\bar{p}$), 
such that an experimental approach can be taken to classify 
events depending on the characteristics of the final state:
\eg gaps or no-gaps, leading protons or neutrons {\it etc}. 

\begin{figure}[tb]
\center{
\begin{tabular}{ll} 
\begin{tabular}{l} 
\epsfig{width=0.22 \columnwidth,file=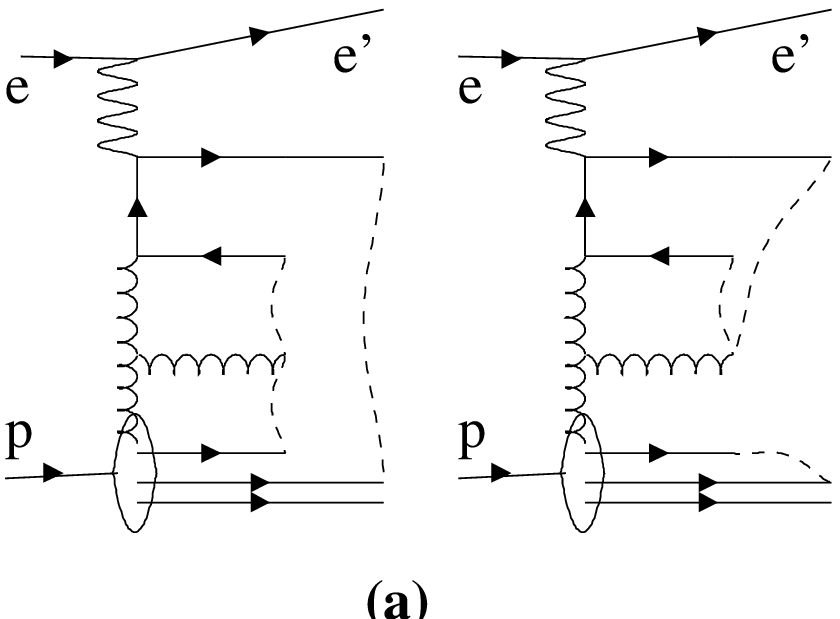, bbllx=0, bblly=24, bburx=121, bbury=178, clip=} \\
\epsfig{width=0.22 \columnwidth,file=sci_diag.eps, bbllx=122, bblly=24, bburx=242, bbury=178, clip=} \\
\end{tabular}
&
\hspace*{10mm}
\epsfig{width=0.36 \columnwidth, file=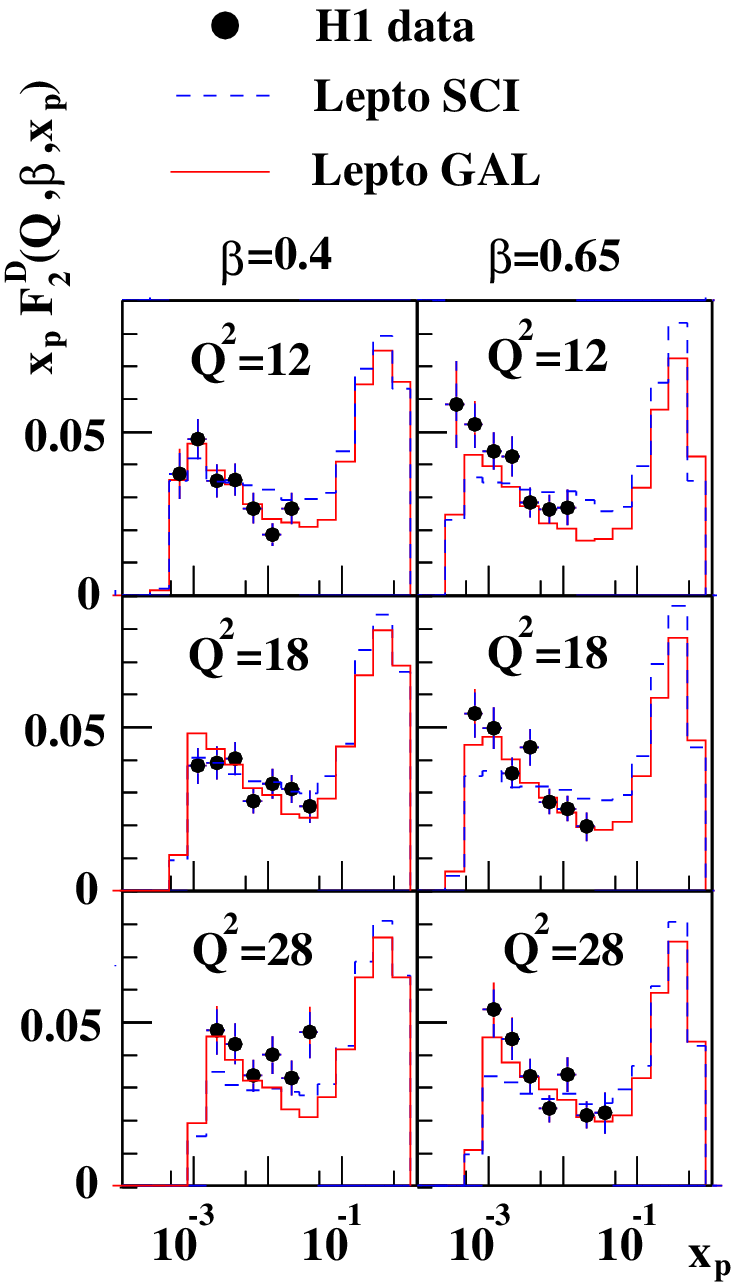, bbllx=0, bblly=161, bburx=212, bbury=369}
\end{tabular}
}
\vspace*{-5mm}
\caption[junk]{Gluon-induced deep inelastic scattering event with examples of 
colour string connection (dashed lines) of partons in conventional Lund model 
\cite{lund}
based on the colour order in pQCD (upper left), and after colour rearrangement through 
SCI or GAL mechanisms (lower left). 
The diffractive structure function $F_2^{D(3)}(x_{\Pma};\beta ,Q^2)$ 
in a selected kinematic range with H1 data \cite{H1}
compared to the SCI and GAL models; excerpt from \cite{GAL-HERA}. 
\label{fig:DIS} \label{fig:f2d3}}
\end{figure}

\begin{figure}[tb]
\center{
\begin{tabular}{ll} 
\begin{tabular}{l} 
\epsfig{width=0.22 \columnwidth,file=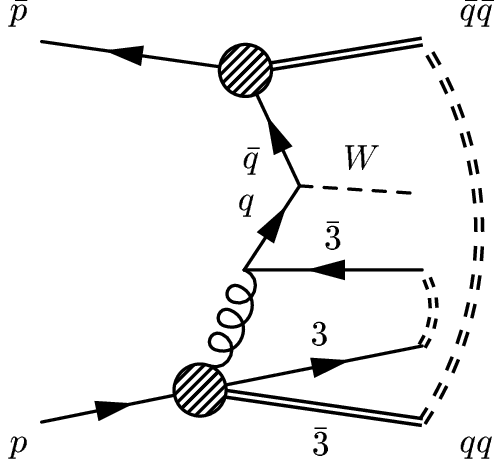} \vspace*{5mm} \\ 
\epsfig{width=0.22 \columnwidth,file=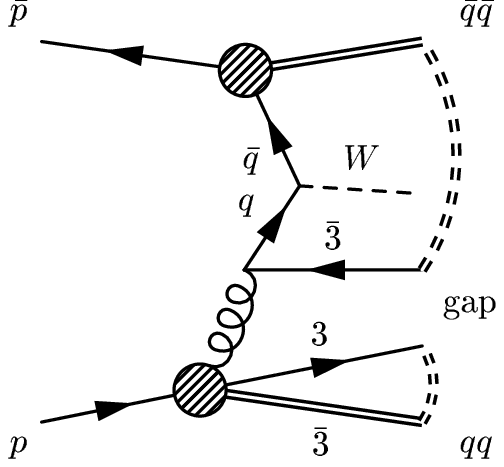} \\ 
\end{tabular}
&
\hspace*{10mm}
\epsfig{width=0.4 \columnwidth, file=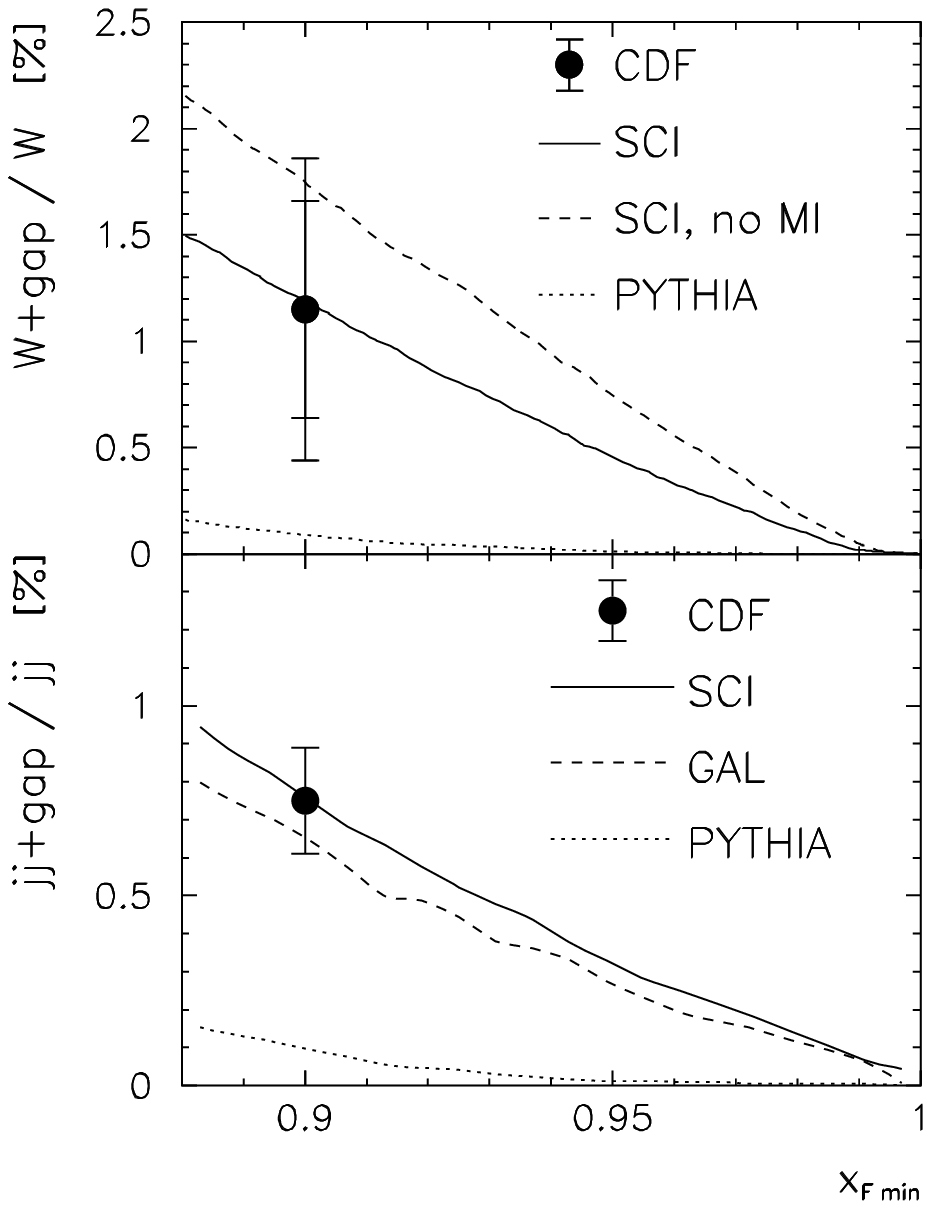, bbllx=0, bblly=200, bburx=283, bbury=397} 
\end{tabular}
}
\caption{$W$ production in $p\bar{p}$ with string topology before 
and after colour rearrangement resulting in a gap. 
Relative rate of $W$ and two-jet ($jj$) events with a rapidity gap 
corresponding to diffraction with a leading proton with minimum $x_F=0.9$
in Tevatron data \cite{CDF-W+dijets}
and in the SCI and GAL models. Sensitivity to multiparton 
interactions (MI) and results from standard {\sc Pythia} without 
soft colour exchanges is also shown \cite{SCI-W-jets}.
}
\label{fig:Wsci}
\end{figure}

The basic assumption of the models is that variations in the topology of the
confining colour  force fields (strings \cite{lund}) lead to different hadronic final
states after hadronisation, as illustrated in Figs.\ \ref{fig:DIS} and 
\ref{fig:Wsci}. The pQCD interaction gives a set of partons with a specific 
colour order. However, this order may change due to soft, 
non-perturbative interactions. 

In the {\em soft colour interaction} (SCI) model \cite{SCI} it is assumed that
colour-anticolour, corresponding to non-perturbative gluons, can be  exchanged
between partons and remnants emerging from a hard scattering.  This can be
viewed as the partons interacting softly with the colour medium  of the proton
as they propagate through it,  which should be a natural part of the process in
which `bare' perturbative  partons are `dressed' into non-perturbative ones and
the confining colour  flux tube between them is formed.  The hard parton level
interactions are given by standard perturbative matrix  elements and parton
showers, which are not altered by softer non-perturbative effects. The unknown
probability to exchange a soft gluon between  parton pairs is
given by a phenomenological parameter $R$, which is the only free parameter of
the model.  With $R=0.5$ one obtains \cite{GAL-HERA} the correct rate of 
rapidity gap events
observed at  HERA and a quite decent description of the measured diffractive
structure  function (Fig.~\ref{fig:f2d3}).

Leading neutrons are also obtained in agreement with experimental
measurements \cite{leading-pn}. In the Regge approach pomeron exchange would 
be used for diffraction, pion exchange added to get leading neutrons and still
other exchanges should be added for completeness. 
The SCI model provides a simpler description.

Applying the same SCI model to hard $p\bar{p}$ collisions 
one obtains production of $W$ and jets in association with rapidity gaps.  
As shown in Fig.~\ref{fig:Wsci}, the model reproduces the rates observed 
at the Tevatron using the same $R$-value as obtained from gaps at HERA. 
This is in contrast to the Pomeron model which, when tuned to HERA gap events, 
gives a factor $\sim 6$ too large rate at the Tevatron 
\cite{StCroix,SCI-W-jets}. 

\begin{figure}[tb]
\vspace*{-10mm}
\center{
\epsfig{file=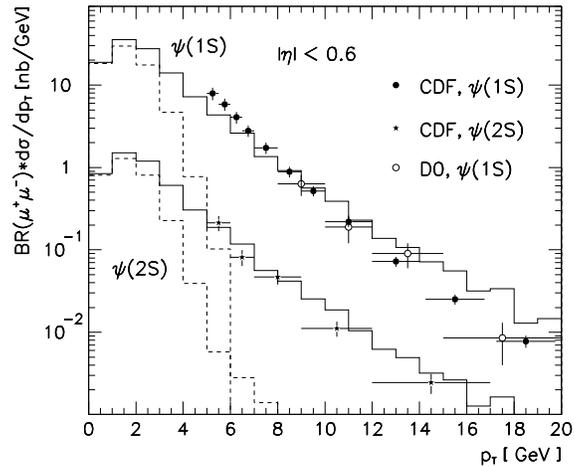,width=0.5 \columnwidth}
}
\vspace*{-10mm}
\caption{Transverse momentum spectra of prompt charmonium 
($J/\psi$ and $\psi ' $) 
in $p\bar{p}$ collisions at the Tevatron energy. 
Data from CDF % \cite{CDF-psi}
and D0        % \cite{D0-psi} 
compared to results from the SCI model (histograms) 
applied to $c\bar{c}$ production from leading order ($\alpha_s^2$) matrix 
elements (dashed lines) and with inclusion of higher order contributions 
calculated in the parton shower approach (full lines). 
From \mbox{\cite{SCI-psi}}.  
}
\label{fig:onium}
\vspace*{-5mm}
\end{figure}

SCI does not only lead to rapidity gaps, but also to other striking effects. 
It reproduces (Fig.~\ref{fig:onium})
the observed rate of high-$p_{\perp}$ charmonium 
and bottomonium at the Tevatron, which are factors of 10 larger than 
predictions based on conventional pQCD. This is accomplished by the 
change of the colour charge of a $Q\bar{Q}$ pair (\eg from a gluon)
from octet to singlet. A quarkonium state can then be formed  using a simple 
model for the division of the cross-section below the threshold for open heavy 
flavour production onto different quarkonium states \cite{SCI-psi}. 

An alternative to SCI is 
the newly developed {\em generalised area law} (GAL) model \cite{GAL} which,
based on a generalisation of the area law suppression $e^{-bA}$ with 
$A$ the area swept out by the string in energy-momentum space, 
gives modified colour string topologies through string reinteractions. 
The probability $P=R_0[1-exp(-b\Delta A)]$ 
for two strings pieces to interact depends on the area difference
$\Delta A$ which is gained by the string rearrangement. This
favours making `shorter' strings, \eg with gaps, whereas making `longer',
`zig-zag' shaped strings is suppressed. The
fixed probability $R$ in SCI is thus  replaced by a dynamical one, where the
parameter $R_0=0.1$ is chosen to reproduce the HERA gap event rate in a 
simultaneous fit to data from $e^+e^-$ annihilation at the $Z^0$-peak. The
resulting diffractive structure function  compares very well with HERA data
(Fig.~\ref{fig:f2d3}). The GAL model also improves the description of
non-diffractive HERA data \cite{GAL-HERA}.

The GAL model can also be applied to $p\bar{p}$ to obtain diffractive $W$ 
and jet production through string rearrangements like in Fig.~\ref{fig:Wsci}.
The observed rates are reproduced quite well (Fig.~\ref{fig:Wsci}). 
However, the treatment of the `underlying event', which is a notorious 
problem in hadron-hadron scattering, introduces a larger uncertainty than 
for the SCI model \cite{SCI-W-jets}. 

In conclusion, our  models for non-perturbative QCD dynamics in terms of 
varying colour string topologies give a satisfactory unified description 
of several phenomena in different hadronic final states. 
This should contribute to a better understanding of non-perturbative QCD
interactions.

%\vspace*{-2mm}

\end{document}